\documentclass[12pt,reqno,a4paper]{amsart}
\usepackage{amsmath,amssymb,dsfont,graphicx,cite,latexsym,epsf,cancel}
\usepackage[colorlinks=false]{hyperref}
\def\beq{\begin{equation}}
\def\eeq{\end{equation}}
\def\bea{\begin{eqnarray}}
\def\eea{\end{eqnarray}}
\newtheorem{theorem}{Theorem}

\newtheorem{rem}{Remark}

\makeatletter
\expandafter\let\expandafter
\reset@font\csname reset@font\endcsname
\def\subeqnarray{\arraycolsep1pt
   \def\@eqnnum\stepcounter##1{\stepcounter{subequation}
       {\reset@font\rm(\theequation\alph{subequation})}}
\jot5mm     \eqnarray}

\makeatother

\def\ep{\varepsilon}
\def\epsilon{\varepsilon}
\def\t{\widetilde}
\def\tilde{\widetilde}
\def\nn{\nonumber}
\def\endpf{\hfill$\blacksquare$\medskip}
\newbox\meibox
\def\placeunder#1#2#3#4{\setbox\meibox%
\vbox{\hbox{\hskip#4$\hphantom{#2}$}\hbox{$\hphantom{#1}$}}%
\vtop{\baselineskip=0pt\lineskiplimit=\baselineskip%
\lineskip=#3\hbox to \wd\meibox{\hfil\hskip#4$#2$\hfil}%
\hbox to \wd\meibox{\hfil$#1$\hfil}}}
\def\intprod{\mathbin{\hbox to 6pt{%
                 \vrule height0.4pt width5pt depth0pt
                 \kern-.4pt
                 \vrule height6pt width0.4pt depth0pt\hss}}}
\begin{document}
\title[S.~Kovalevskaya system, its generalization and discretization]
{S.~Kovalevskaya system, \\ its generalization and discretization}

\author{Matteo Petrera \and Yuri B. Suris }

\thanks{E-mail: {\tt  petrera@math.tu-berlin.de, suris@math.tu-berlin.de}}

\maketitle

\begin{center}
{\footnotesize{
Institut f\"ur Mathematik, MA 7-2,\\
Technische Universit\"at Berlin, Str. des 17. Juni 136,
10623 Berlin, Germany
}}
\end{center}


\begin{abstract}
We consider an integrable three-dimensional system of ordinary differential equations introduced by S.V. Kovalevskaya
in a letter to G. Mittag-Leffler. We prove its isomorphism with the three-dimensional Euler top, and propose two integrable
discretizations for it. Then we present an integrable generalization of the Kovalevskaya system, and study the problem of integrable discretization for this generalized system.
\end{abstract}

\section{Introduction}

The problem of finding integrable discretizations for integrable systems of ordinary differential equations was studied a lot in recent years, see \cite{S}. Depending on the integrability attributes one wants to respect, different approaches can be proposed and sometimes different results can be achieved for a given integrable system. One common feature of almost all integrable discretizations known to date is that equations of motion of these maps appear to be rather nontrivial deformations of the original differential equations. These deformations can only be found after running a sometimes involved machine of some systematic approach, or, alternatively, by a guesswork based on an extensive experience and expert knowledge.

We are aware of only one approach for which the situation is different. This approach, due to R. Hirota and K. Kimura \cite{HK}, is applicable to any vector field with a quadratic dependence on coordinates. It consists in replacing the derivatives by the first differences and the quadratic terms in the vector fields by the corresponding bilinear terms (with respect to the values of coordinates in two subsequent instances of the discrete time). Thus, a derivation of a Hirota-Kimura discretization for a system is quite straightforward and algorithmic. However, there are no theoretical reasons for Hirota-Kimura discretizations to preserve integrability. Nevertheless, very remarkably, it turns out that very often they do! An extensive experimental and theoretical study of this phenomenon is contained in our recent works \cite{PS,PPS1,PPS2,HP}. These papers contain a big collection of novel integrable birational maps, whose integrability is expected to be related to some deep phenomena in algebraic geometry.

The goal of the present work is to add several new instances to this collection, with the hope that this will facilitate the development of the still lacking theoretical explanation of these exciting observations.

\section{Kovalevskaya system and its relation to the Euler top}

In a letter to G. Mittag-Leffler from December 1884 \cite[p. 80--82]{K}, S. Kovalevskaya discussed a class of systems of ordinary differential equations with quadratic vector fields,
\beq \label{kov gen}
\dot{y}_i=y_i\sum_{j=1}^3a_{ij}y_j,\qquad i=1,2,3,
\eeq
where $a_{ij}$ are parameters.
She mentioned that under the condition $a_{12}a_{23}a_{31}=a_{13}a_{32}a_{21}$, a general solution of such a system depends on three arbitrary parameters, and its only singularities in the finite part of the complex plane of the independent variable are first order poles. In the modern theory, this property is called ``Painlev\'e property" and is associated (somewhat vaguely) with integrability. As a particular case of this class, she mentioned the following system:
\beq \label{kow}
\left\{ \begin{array}{l}
\dot y_1 = y_1 (-y_1 + y_2 + y_3),\vspace{.1truecm} \\
\dot y_2 = y_2 (-y_2 + y_3 + y_1) , \vspace{.1truecm} \\
\dot y_3 = y_3 (-y_3 + y_1 + y_2).
 \end{array} \right.
\eeq
This system admits three conserved quantities:
\beq
\label{qq3}
K_{23}=y_1(y_2-y_3), \qquad K_{31}=y_2(y_3-y_1), \qquad K_{12}=y_3(y_1-y_2).
\eeq
They are linearly dependent, $K_{23}+K_{31}+K_{12}=0$, and any two of them are functionally independent. In her letter, S. Kovalevskaya stated that system (\ref{kow}) can be integrated in terms of elliptic functions and suggested that the study of the class of systems (\ref{kov gen}) possessing the Painlev\'e property might lead to a better understanding of general systems with quadratic vector fields. Although this opinion is probably too optimistic, it is definitely useful to have a closer look at system (\ref{kow}) and its integrable generalizations.

What S. Kovalevskaya did not mention is a simple (and even two-fold) relation of system (\ref{kow}) to a much more famous integrable system, the Euler top (see, e.g., \cite{RSTS}):
\beq\label{euler}
\left\{ \begin{array}{l}
\dot x_1 = x_2 x_3 ,\vspace{.1truecm} \\
\dot x_2 = x_3 x_1  , \vspace{.1truecm} \\
\dot x_3 = x_1 x_2 .
 \end{array} \right.
\eeq
One can bring the latter system to the form $\dot x_i=\alpha_ix_jx_k$, involving arbitrary real parameters $\alpha_i$, by a simple scaling transformation $x_i\mapsto \beta_ix_i$, where, however, some of $\beta_i$ may be imaginary, depending on the signs of $\alpha_i$. Here and below we often use $(i,j,k)$ to denote a cyclic permutation of (1,2,3).

System (\ref{euler}) admits three integrals of motion
$$
E_{ij} = x_i^2-x_j^2,
$$
only two of which are functionally independent due to $E_{23}+E_{31}+E_{12}=0$. It is easy to verify that both the following changes of variables, a linear one,
\begin{equation} \label{is2}
 y_i = \frac12 (x_j+x_k) \qquad \Leftrightarrow \qquad x_i = -y_i + y_j+y_k,
\end{equation}
and a nonlinear one,
\begin{equation}\label{isrt}
    y_i = \frac{x_j x_k}{x_i} \qquad \Leftrightarrow \qquad x_i = \sqrt{y_j y_k},
\end{equation}
map (\ref{euler}) to (\ref{kow}). Indeed, for the change of variables (\ref{is2}) we compute:
\beq \nonumber
\dot y_i = \frac12 (\dot x_j + \dot x_k)= \frac{1}{2} x_i(x_j+x_k)
= y_i(-y_i + y_j+y_k),
\eeq
while for the change of variables (\ref{isrt}) the corresponding computation gives:
\beq \nonumber
\frac{\dot y_i}{y_i}=\frac{\dot x_j}{x_j} + \frac{\dot x_k}{x_k} - \frac{\dot x_i}{x_i}= \frac{x_i x_k}{x_j} +  \frac{x_i x_j}{x_k} - \frac{x_j x_k}{x_i} = -y_i + y_j+y_k.
\eeq
One easily sees that the conserved quantities $K_{ij}$ of (\ref{kow}) are obtained from the corresponding conserved quantities $E_{ij}$ of (\ref{euler}) by either of the changes of variables.

\section{Two integrable discretizations of the  Kovalevskaya system}

Any system of ordinary differential equations with a quadratic vector field can be discretized via the Hirota-Kimura (HK) approach, which consists of the replacement of the derivatives $\dot{x}_j$ by the differences $(\t x_j-x_j)/\epsilon$ and the quadratic expressions like $x_jx_k$ or $x_j^2$ by their respective bilinear counterparts, $x_j\t x_k+\t x_jx_k$, resp. $2x_j\t x_j$. For a function $x:\epsilon\mathbb Z\to\mathbb R$, we use the abbreviation $x$ for $x(t)$ and $\t x$ for $x(t+\epsilon)$, where $\epsilon$ is the discrete time-step. We refer to \cite{PS,PPS1,PPS2,HP}
for our recent investigations on HK-type discretizations.

The HK-discretization of the Euler top (\ref{euler})  is described by the following discrete equations of motion \cite{HK}:
\beq
\left\{ \begin{array}{l}
\t x_1 - x_1 = \ep (\t x_2 x_3 + x_2 \t x_3) ,\vspace{.1truecm} \\
\t x_2 - x_2 = \ep (\t x_3 x_1 + x_3 \t x_1) , \vspace{.1truecm} \\
\t x_3 - x_3 = \ep (\t x_1 x_2 + x_1 \t x_2).
 \end{array} \right. \label{deuler}
\eeq
These equations define a birational map $(\t x_1,\t x_2,\t x_3)=f(x_1,x_2,x_3,\epsilon)$, reversible because of $f^{-1}(x,\epsilon)=f(x,-\epsilon)$. Its explicit form is:
\beq
\label{deuler expl}
\widetilde{x}_i=\dfrac{x_i+2\epsilon x_jx_k+
\epsilon^2x_i(-x_i^2+x_j^2+x_k^2)}
{1-\epsilon^2(x_1^2+x_2^2+x_3^2)
-2\epsilon^3x_1x_2x_3}.
\eeq
As proven in \cite{PS}, this map is bi-Hamiltonian and completely integrable. Its orbits are given in terms of elliptic functions, and it admits conserved quantities
\beq \nn
E_{mn} (\ep)=\frac{  x_m^2 - x_n^2}{1 - \ep^2   x_j^2},
\eeq
with arbitrary $m,n,j=1,2,3$. (This notation is incomplete as it does not refer to $j$; only the indices surving in the continuous limit $\ep\to 0$ are explicitly referred to.) Among the functions $E_{mn} (\ep)$, there are two functionally independent ones. Moreover, the map (\ref{deuler expl})  has an invariant volume form
\beq \nn
\omega= \frac{dx_1\wedge dx_2 \wedge dx_3}{\phi(x)},
\eeq
where $\phi(x)$ is any of the functions $(1 - \ep^2  x_j^2)^2$. Note that the ratio of any two such functions is a conserved quantity.

It turns out that the HK-discrete Euler top (\ref{deuler}) can be transformed into birational integrable discretizations of the Kovalevskaya system (\ref{kow}) by any of the changes of variables (\ref{is2}) and (\ref{isrt}). However, the resulting maps are different.

Linear changes of variables preserve the class of HK-discretizations. Therefore, the change of variables (\ref{is2}) maps the HK-discrete Euler top onto the HK-discrete Kovalevskaya system:
\beq \label{dkow}
\left\{ \begin{array}{l}
\t y_1 - y_1 = \ep  (-2 \t y_1 y_1 + \t y_1 y_2 + y_1 \t y_2+  \t y_1 y_3 + y_1 \t y_3) ,\vspace{.1truecm} \\
\t y_2 - y_2 =\ep  (-2 \t y_2 y_2 + \t y_2 y_3 + y_2 \t y_3+  \t y_2 y_1 + y_2 \t y_1) , \vspace{.1truecm} \\
\t y_3 - y_3 = \ep  (-2 \t y_3 y_3 + \t y_3 y_1 + y_3 \t y_1+  \t y_3 y_2 + y_3 \t y_2).
 \end{array} \right.
\eeq

The properties of the HK-discrete Euler top are then translated as follows.

\begin{theorem}\label{th: dkow}
${}$
\begin{enumerate}

\item The map (\ref{dkow}) admits the following conserved quantities:
\beq
K_{mn}(\ep) = \frac{K_{mn}}{1 - \ep^2 (y_i - y_j + y_k)^2}, \label{ddefg}
\eeq
where $K_{mn}$, with arbitrary and distinct $m,n=1,2,3$ are defined in (\ref{qq3}) and $(i,j,k)$
is any permutation of $(1,2,3)$. Among functions (\ref{ddefg}) only
two are functionally independent.
\\
\item The map (\ref{dkow}) admits an invariant volume form
\beq \nn
\omega= \frac{dy_1\wedge dy_2 \wedge dy_3}{\psi(y)},
\eeq
where $\psi(y)$ is any of the functions $(1 - \ep^2 (y_i-y_j+y_k)^2)^2$.

\end{enumerate}
\end{theorem}

Now we apply the nonlinear change of variables (\ref{isrt}) to the HK-discretization of the Euler top
(\ref{deuler}). A remarkable feature of the map (\ref{deuler expl}), written as
$$
\frac{\widetilde{x}_i}{x_i}=\dfrac{1+2\epsilon (x_jx_k/x_i)+
\epsilon^2(-x_i^2+x_j^2+x_k^2)}
{1-\epsilon^2(x_1^2+x_2^2+x_3^2)
-2\epsilon^3x_1x_2x_3},
$$
is that the r.h.s~of the latter formula  depends rationally on combinations of $x_i$ of the form
$
x_jx_k/x_i=y_i,\, x_i^2=y_jy_k,$ and $ x_1x_2x_3=y_1y_2y_3,
$
which do not involve square roots $\sqrt{y_j}$. As a consequence, the pull-back of the HK-discrete Euler top under change of variables (\ref{isrt}) is birational:
\beq\label{pp}
\t y_i =y_i  \frac{\big(1+ 2 \ep y_j + \ep^2 (y_i y_j + y_j y_k - y_k y_i )\big) \big(1+ 2 \ep y_k + \ep^2 (y_j y_k + y_k y_i - y_i y_j )\big)}
{\big(1+ 2 \ep y_i + \ep^2 (y_k y_i + y_i y_j - y_jy_k )\big) \big(1- \ep^2 (y_1 y_2 + y_2 y_3 + y_3 y_1 ) -
2 \ep^3 y_1y_2 y_3\big)}.
\eeq

Before translating properties of the HK-discrete Euler top into properties of map (\ref{pp}), we mention a result of our recent work \cite{PS2}, according to which the HK-discretization of the Euler top (\ref{deuler expl}) is the second iterate of another map which is nothing but the cosine law for spherical triangles in $\mathbb{S}^2$. Indeed, the cosine law can be written as 
\[
\overline x_{i}=  \frac{x_{i} +x_{j}x_{k}}
{\sqrt{1-x_j^2}\sqrt{1-x_k^2}},
\]
where $x_i=\cos\alpha_i$, $\overline x_i=\cos\ell_i$, with $\alpha_i$ and $\ell_i$ being angles and sides, respectively, of a spherical triangle. The small parameter $\ep$ can be inserted by a scaling transformation $x_i \mapsto \ep x_i$, $\overline x_i \mapsto \ep \overline x_i$, which makes out of the latter map a ``new'' discretization of the Euler top:
\beq\label{rt}
\overline x_{i}=  \frac{x_{i} +\epsilon x_{j}x_{k}}
{\sqrt{1-\ep^2 x_j^2}\sqrt{1-\ep^2 x_k^2}}.
\eeq
The second iterate of (\ref{rt}), where defined, is exactly the birational map (\ref{deuler expl}).

Now, performing the nonlinear change of variables (\ref{isrt}) in (\ref{rt}), we easily get:
\beq \label{jfg}
\overline y_{i}=y_i\frac{(1+ \ep y_j)(1+\ep y_k)}{(1+\ep y_i)(1-\ep^2 y_j y_k)},
\eeq
which, remarkably, is birational. The second iterate of (\ref{jfg})  is, of course, the map (\ref{pp}). One can say that the latter map admits a birational ``square root" map. 

The properties of the HK-discrete Euler top are translated under (\ref{isrt}) as follows.

\begin{theorem}\label{THv}
${}$
\begin{enumerate}

\item
The map (\ref{jfg}) and its second iterate (\ref{pp}) admit the following conserved quantities:
\beq
\label{intk2}
K_{mn}(\ep) = \frac{K_{mn}}{1-\ep^2 y_iy_j},
\eeq
where $(m,n)$ and $(i,j)$ are two arbitrary pairs of distinct numbers from $\{1,2,3\}$, and $K_{mn}$ are defined in (\ref{qq3}). Among functions (\ref{intk2}) there are two functionally independent ones.
\\
\item The map (\ref{jfg}) and its second iterate (\ref{pp}) admit an invariant volume form
\beq \nn
\omega= \frac{dy_1\wedge dy_2 \wedge dy_3}{\psi(y)},
\eeq
where $\psi(y)$ is any of the functions $(1-\ep^2 y_iy_j)^2$.

\end{enumerate}
\end{theorem}

One sees immediately that the ratio of any two functions $K_{mn}$ from (\ref{qq3}) serves as an integral of motion for the both discretizations (\ref{deuler expl}) and (\ref{jfg}). Moreover, they also share an invariant
volume form with an $\ep$-independent density $1/K_{mn}^2$.

\section{Generalized Kovalevskaya system}

The main subject of investigation in the present paper is the following system with a quadratic vector field, which is a generalization of the Kovalevskaya system (\ref{kow}):
\beq\label{kk3}
\dot{y}_i=y_i(-2 y_i + s) ,\qquad
1\leq i\leq N,
\eeq
with
\beq\label{sssss}
s=\sum_{j=1}^N y_j.
\eeq
System (\ref{kk3}) was proposed in \cite[p. 55--56]{BM}, where, however, only one new result was established for this system, namely existence of three polynomial conserved quantities for $N=4$:
\beq \label{bm}
P_1 = (y_1 - y_2)(y_3-y_4), \quad
P_2 = (y_1 - y_3)(y_2-y_4), \quad
P_3 = (y_1 - y_4)(y_2-y_3).
\eeq
Actually, only two of them are functionally independent, because of $P_1-P_2+P_3=0$.

One can relate system (\ref{kk3}) to an elegant generalization of the Euler top proposed in \cite{F}:
\beq\label{hb}
\dot x_i = \prod_{j \neq i} x_j, \qquad 1\le i\le N.
\eeq
System (\ref{hb}) preserves the phase volume (since its vector field is obviously divergence-free) and admits a set of conserved quantities
\beq\label{ll}
E_{ij} = x_i^2-x_j^2.
\eeq
Since one can find $N-1$ functionally independent conserved quantities among $E_{ij}$, system (\ref{hb}) is superintegrable. Moreover, it can be explicitly integrated as follows: use the integrals to express $N-1$ variables, say $x_j$ with $2 \leq j \leq N$, in terms of one variable, say $x_1$, leading to a differential equation for $x_1$:
$$
(\dot x_1)^2=  \prod_{j=2}^N (x_1^2 + E_{j1}).
$$
Thus, integration is performed through the inversion of a hyperelliptic integral.

Systems (\ref{kk3})  and (\ref{hb}) are related by a change of variables which is a direct generalization of the nonlinear change of variables (\ref{isrt}):
\beq\label{gKF}
y_i = \frac{1}{x_i}  \prod_{j \neq i} x_j \qquad\Leftrightarrow\qquad
x_i^2= \frac{1}{y_i} \left( \prod_{j=1}^N y_j\right)^{\frac{1}{N-2}} .
\eeq
This is shown by a direct computation:
$$
\frac{\dot y_i}{y_i}=-\frac{\dot x_i}{x_i}+\sum_{j\neq i}\frac{\dot x_j}{x_j}=-y_i+\sum_{j\neq i} y_j=-2y_i+s.
$$

The properties of system (\ref{hb}) are translated into properties of system (\ref{kk3}) as follows.

\begin{theorem}\label{th: kov}
${}$
\begin{enumerate}

\item The functions
\beq \label{uu}
K_{ij}= \frac{y_i - y_j}{y_iy_j} \left(\prod_{k=1}^N y_k\right)^{\frac{1}{N-2}}
\eeq
are conserved quantities of  system (\ref{kk3}). Among functions (\ref{uu}), there are $N-1$ functionally independent ones.
\\
\item System (\ref{kk3}) admits an invariant volume form
\beq
\omega = \frac{dy_1 \wedge \cdots \wedge dy_N}{\phi(y)}, \qquad \phi(y)=\left(\prod_{k=1}^N y_k \right)^{\frac{N-3}{N-2}}. \label{miss}
\eeq
\end{enumerate}
\end{theorem}

\noindent{\it{Proof.}} The first statement is an immediate corollary of (\ref{ll}) and (\ref{gKF}). The second follows from an easily verified formula
\[
\det\left(\frac{\partial y}{\partial x}\right)=\left(\prod_{i=1}^Nx_i\right)^{N-3}=
\left(\prod_{i=1}^N y_i\right)^{\frac{N-3}{N-2}}. \put(110,0){$\blacksquare$}
\]

Thus, system (\ref{kk3}) is superintegrable for any $N$. In particular, for $N=4$ one has three independent integrals among
\beq\label{qq4}
K_{ij}=(y_i-y_j)\sqrt{\frac{y_ky_\ell}{y_iy_j}},
\eeq
where $(i,j,k,\ell)$ is any permutation of $(1,2,3,4)$. Integrals (\ref{bm}) are obtained from those as follows: $P_1=K_{12} K_{34}, \, P_2=K_{13} K_{24}, \, P_3=K_{14} K_{23}.$

We remark that system (\ref{kk3}) admits a further integrable generalization:
\beq
\dot{y}_i=y_i(-\alpha y_i + s) ,\qquad
1\leq i\leq N, \label{kk3bis}
\eeq
with an arbitrary real $\alpha\neq N$. A direct computation shows that (\ref{kk3bis}) has $N-1$ functionally independent conserved quantities, contained in the set of functions
\beq
K_{ij}= \frac{y_i - y_j}{y_iy_j} \left(\prod_{k=1}^N y_k\right)^{\frac{1}{N-\alpha}}, \label{rtghh}
\eeq
and an invariant volume form
$$
\omega =\frac{dy_1 \wedge \cdots \wedge dy_N}{ \phi(y)}, \qquad
\phi(y)=\left(\prod_{k=1}^N y_k \right)^{\frac{N+1-2\alpha}{N-\alpha}}.
$$
Moreover, one can establish the existence of $N-2$ independent integrals of motion for a still more general system, namely for (\ref{kk3bis}) with any symmetric function $s$, not necessarily given by (\ref{sssss}). Indeed, setting
$H_{ij}=(y_i-y_j)/(y_i y_j),$ we immediately derive from (\ref{kk3bis}):
$$
\dot H_{ij}= \frac{\dot{y}_i}{y_i^2}-\frac{\dot{y}_j}{y_j^2}= s\left(\frac{1}{y_i} -\frac{1}{y_j} \right)=-s H_{ij}.
$$
Therefore, the functions $H_{ij}/H_{k\ell}=K_{ij}/K_{k\ell}$ are conserved quantities irrespectively of $s$.

\section{Hirota-Kimura discretization of the generalized Kovalevskaya system}

Unlike in the  case $N=3$, the relation to the top-like system (\ref{hb}) does not help to find integrable discretizations for (\ref{kk3}), since we are not aware of any decent integrable discretization of the former system. On the contrary, since system (\ref{kk3}) has a quadratic vector field, we are in a position to apply a general Hirota-Kimura discretization scheme to it. The resulting discrete equations of motion read:
\beq \label{map}
\tilde y_i - y_i = \ep \left(-4  y_i \tilde y_i+y_i \t s+\tilde y_is \right), \qquad 1 \leq i \leq N,
\eeq
where $s$ is defined as in (\ref{sssss}).
As usual for HK-discretizations, equations of motion (\ref{map}) can be solved for $\t y$, yielding the rational map
\beq\label{map expl}
\tilde y = f(y,\ep)=A^{-1}(y, \ep) y, \qquad y = (y_1, \dots ,y_N)^{\rm{T}},
\eeq
with
$$
A(y,\ep)=\begin{pmatrix}
 1-\epsilon (-3 y_1 + s) & -\epsilon y_1   &  \ldots &  -\epsilon y_1 \\
-  \epsilon y_2 & 1-\epsilon (-3 y_2 + s) & \ldots &  -\epsilon y_2\\
 \ldots & \ldots & \ldots  &  \ldots  \\
 -\epsilon y_{N-1} & -\epsilon y_{N-1}  & \ldots & -\epsilon y_{N-1}\\
 -\epsilon y_N & -\epsilon y_N & \ldots  & 1- \ep (-3 y_{N} + s)
\end{pmatrix}.
$$
This map is actually birational, due to the reversibility property $f^{-1}(y,\ep)=f(y,-\ep)$.

\begin{theorem}
Map (\ref{map expl}) can be represented in the following explicit form:
\beq \label{ss}
\t y_i=\frac{1}{S(y,\ep)}\, \frac{y_i}{d_i}, \qquad 1 \leq i \leq N,
\eeq
where
\beq\label{defd}
d_i  = 1- \ep ( -4 y_i + s),
\eeq
and
\beq \label{defS}
S(y, \ep)=  1 - \ep \, \sum_{j=1}^N \frac{y_j}{d_j}.
\eeq
\end{theorem}

\noindent {\it Proof.} The matrix $A(y,\epsilon)$ may be written as
$$
A(y,\ep) = D-\ep ye^{\rm T},
\qquad
D = {\rm{diag}} \left(d_1, \dots, d_N \right),
$$
where  $e= (1, \dots, 1)^{\rm T}$. Therefore,
\bea
\tilde y &=&  A^{-1}(y,\ep)y \ = \left(\mathds{1}-\ep D^{-1}ye^{\rm T}\right)^{-1} D^{-1}y \nn \\
&=&\left(\mathds{1}+\ep\frac{D^{-1}ye^{\rm T}}{1-\ep\langle\,  D^{-1}y,e\, \rangle}\right)D^{-1}y\nonumber\\
&=& D^{-1}y\left(1+\ep\,\frac{\langle \, D^{-1}y,e\, \rangle}{1-\ep\langle\,  D^{-1}y,e\, \rangle}\right)= \frac{D^{-1}y}{S(y,\ep)} \nonumber,
\eea
where $S=1-\ep\langle D^{-1}y,e\rangle$. This coincides with (\ref{ss}). \endpf

\begin{theorem}
The following formulas hold true:
\beq \label{sd}
S(y,\ep)=\frac{1}{1+\ep\t s}\,, \qquad S(\t y,-\ep)=\frac{1}{1-\ep s}.
\eeq
\end{theorem}

\noindent {\it Proof.}
From (\ref{ss}) and  (\ref{defS}) we derive:
\[
S(y,\ep)=1-\ep\sum_{j=1}^NS(y,\ep)\t y_j,
\]
which gives $S(y,\ep)(1+\ep\t s)=1$. Similarly for the second formula.
\endpf

\begin{theorem}
${}$
\begin{enumerate}

\item
The functions
\beq \label{444}
\frac{K_{ij}}{K_{k\ell}}=\frac{y_i-y_j}{y_iy_j}\;\frac{y_ky_\ell}{y_k-y_\ell},
\qquad 1\leq i,j,k,\ell\leq N,
\eeq
are conserved quantities of the map (\ref{map}). Among them, there are $N-2$ functionally independent ones.
\\
\item
The map (\ref{map}) admits an invariant volume form
\beq
\omega= \frac{dy_1\wedge \cdots \wedge dy_N}{\psi(y)},\nonumber
\eeq
where $\psi(y)$ is any  of the functions
\beq \label{den}
\psi(y) = \left( \frac{y_i-y_j}{y_iy_j}\right)^{N-1}\left(\prod_{k=1}^Ny_k\right)^2.
\eeq
\end{enumerate}
\label{THr2}
\end{theorem}

\noindent {\it Proof.} (1) Equations of motion (\ref{map}) are equivalent to
$$
\frac{1+\ep \t s}{\t y_i}-\frac{1-\ep s}{ y_i}= 4 \ep.
$$
Therefore,
$$
(1+\ep \t s) \left(\frac{1}{\t y_i}-\frac{1}{\t y_j}\right)=(1-\ep s) \left(\frac{1}{ y_i}- \frac{1}{ y_j}\right),
$$
or
\beq \label{qq}
\frac{\t y_i-\t y_j}{\t y_i\t y_j}\;\frac{y_iy_j}{y_i-y_j}=\frac{1-\ep s}{1+\ep\t s}\,.
\eeq
The r.h.s.~of this formula does not depend on indices $i,j$, which proves the first statement of the Theorem.

(2) We start with the following relation, which holds true for any HK-discretization:
$$
\det \left(\frac{\partial\t y}{\partial y} \right)= \frac{\det A(\t y,-\ep)}{\det A(y,\ep)}.
$$
Determinants on the r.h.s~are easily computed:
\bea
\det  A(y,\ep)&=&\det (D)\det (\mathds{1}-\ep D^{-1}ye^{\rm T})\nn \\
&=&S(y,\ep)\prod_{k =1}^N d_k=\frac{1}{S^{N-1}(y,\ep)} \prod_{k=1}^N \frac{y_k}{\t y_k}.
\nn
\eea
In this computation we used the fact that
$
\det (\mathds{1}-\ep D^{-1}ye^{\rm T}) = 1-\ep \, \langle D^{-1}y, e\rangle =S(y,\ep),
$
as well as (\ref{ss}). Upon using the first of equations in (\ref{sd}), we find:
$$
\det A(y,\ep)=(1+ \ep \t s )^{N-1} \prod_{k=1}^N \frac{y_k}{\t y_k}.
$$
Similarly, we have:
$$
\det A(\t y,- \ep)= (1- \ep s )^{N-1} \prod_{k=1}^N \frac{\t y_k}{y_k}.
$$
As a consequence, we find:
\beq
\det \left(\frac{\partial \t y}{\partial y } \right) = \left( \frac{1 - \ep s}{1 + \ep \t s } \right)^{N-1} \left( \prod_{k=1}^N  \frac{\t y_k}{y_k}\right)^2.\nn
\eeq
With the help of (\ref{qq}) this can be represented as
$$
\det \left(\frac{\partial \t y}{\partial y } \right) = \left( \frac{\t y_i-\t y_j}{\t y_i\t y_i} \;
\frac{y_iy_j}{y_i-y_j}\right)^{N-1}\left( \prod_{k=1}^N \frac{\t y_k}{y_k}\right)^2,
$$
which is equivalent to the second statement.
\endpf

\begin{rem}{\rm{
Note that the ratio of any two of distinct functions $\psi(y)$ in (\ref{den}) is an integral of motion, according to  statement (1) of Theorem \ref{THr2}. Note also that
functions $\psi(y)$ do not depend on $\ep$. Therefore,
they must serve as invariant volume densities for the continuous flow (\ref{kk3}), as well. Indeed,  $\psi(y)=K_{ij}^{N-1}(y)\phi(y)$, where  $K_{ij}$ and $\phi$ are integrals and an invariant volume density for the flow (\ref{kk3}), given respectively in (\ref{uu}) and (\ref{miss}).
\qed}}
\end{rem}

\begin{rem}{\rm{Statement (1) of Theorem \ref{THr2} holds true, exactly in the same form and with the same proof,
 for the HK-discretization of the system (\ref{kk3bis}) with arbitrary $\alpha$. Moreover, the specific form of the function $s$ is also irrelevant for the claim and for the proof, as long as it is a symmetric function of the coordinates. \qed}}
\end{rem}

Now the question is whether the map (\ref{map}) admits the last, $(N-1)$-th, integral of motion. Conserved quantities (\ref{uu}) from Theorem \ref{th: kov} suggest to look for it in the form
$$
K_{ij}(\epsilon)= K_{ij}\, \Phi(y,\epsilon),
$$
with some $\Phi(y,\ep)$ which is an even function of $\ep$. It follows from (\ref{qq}) that the necessary and sufficient condition for this is given by the functional equation
$$
\frac{\Phi(\t y, \epsilon)}{\Phi(y, \epsilon)}=\frac{1+\epsilon\t s}{1-\epsilon s}
\left(\prod_{k=1}^N \frac{ y_k}{\t y_k}\right)^{\frac{1}{N-2}}.
$$
It turns out that a solution of the latter equation can be always found as a formal series
 $$
\Phi(y,\epsilon)=1 +\sum_{k=1}^\infty \epsilon^{2k}\varphi_{2k}(y).
$$
However, it seems that this series converges (and defines a genuine function) only in the cases $N=3,4$.
Thus, only in these cases can we claim that the discrete system (\ref{map}) admits the same number of conserved quantities (2 and 3, respectively) as the corresponding continuous system.

In the case $N=3$, the existence of two independent integrals of motion has been already given
in Theorem \ref{th: dkow} as a consequence of the integrability of the
HK-discretization of the Euler system (\ref{deuler expl}).
We found that
the map (\ref{map}) with $N=3$ admits the following conserved quantities:
$$
K_{mn}(\epsilon)=K_{mn}\,\Phi(y,\epsilon),
\qquad \Phi(y,\ep)=\frac{1}{ 1-\epsilon^2( y_i- y_j+y_k)^2}.
$$
Here $K_{mn}$ are defined in (\ref{qq3}), while $(i,j,k)$ is any permutation of $(1,2,3)$. This set of integrals contains two functionally independent ones.

In the case $N=4$, we have the following result.
\begin{theorem}
The map (\ref{map}) with $N=4$ admits the following conserved quantities:
\beq
K_{mn}(\epsilon)=K_{mn}\,\Phi(y,\epsilon),
\qquad
\Phi(y,\ep)=\frac{1}{ \sqrt{1-\epsilon^2(y_i + y_j - y_k - y_\ell )^2 }}.
\label{hujl}
\eeq
Here $K_{mn}$, with arbitrary $m,n=1,2,3,4$, are defined in (\ref{uu}), while $(i,j,k,\ell)$ is any permutation of $(1,2,3,4)$. Among functions (\ref{hujl}), there are three functionally independent ones.

\end{theorem}

 \noindent {\it Proof.}  We have to prove that
\beq \label{88}
 \left( \frac{\t y_m - \t y_n}{\t y_m \t y_n} \; \frac{y_m y_n}{y_m - y_n} \right)^2
  \frac{ \t y_1 \t y_2 \t y_3 \t y_4}{  y_1 y_2 y_3 y_4 }=
  \frac{1-\epsilon^2(\t y_i+\t y_j-\t y_k-\t y_\ell)^2}{1-\epsilon^2(y_i+y_j-y_k-y_\ell)^2}.
\eeq
As a direct consequence of (\ref{ss}) and (\ref{defd}), we have:
\beq \label{qqq}
\tilde y_i - \tilde y_j = \frac{1- \ep s}{ S(y , \ep)} \, \frac{{y_i - y_j}}{d_i d_j}.
\eeq
This formula, together with (\ref{ss}), yields the following expression for the l.h.s.~of  (\ref{88}):
\beq \label{p1}
 \left( \frac{\t y_m - \t y_n}{\t y_m \t y_n} \;\frac{y_m y_n}{y_m - y_n} \right)^2
 \frac{\t y_1\t y_2\t y_3\t y_4}{y_1 y_2 y_3 y_4}=\frac{(1-\epsilon s)^2}{S^2}\frac{1}{d_1d_2d_3d_4}.
\eeq
To find a proper expression for the r.h.s.~of (\ref{88}), we start with the following formula for $S$:
 \beq \label{se}
 S = \frac{(1-\ep s)}{4}\left(\frac{1}{d_1}+\frac{1}{d_2}+\frac{1}{d_3}+\frac{1}{d_4}\right).
 \eeq
It is obtained upon taking into account  (\ref{defS}) and expressing $y_i=(d_i-(1-\ep s))/(4\ep)$ by virtue of (\ref{defd}). Now, using (\ref{qqq}) to express the differences $\t y_i-\t y_k$ and $\t y_j-\t y_\ell$ and  expressions $y_i-y_k=(d_i-d_k)/(4\epsilon)$ following from  (\ref{defd}), we have:
\[
  1-\epsilon^2(\t y_i + \t y_j -\t  y_k - \t y_\ell)^2 = 1- \frac{(1-\ep s)^2}{16S^2} \left(\frac{1}{d_i}-\frac{1}{d_k}+\frac{1}{d_j}-\frac{1}{d_\ell}\right)^2.
\]
Upon using  (\ref{se}), we find:
\[
  1-\epsilon^2(\t y_i+\t y_j-\t y_k-\t y_\ell)^2=\frac{(1-\ep s)^2}{4S^2} \left(\frac{1}{d_i}+\frac{1}{d_j}\right)\left(\frac{1}{d_k}+\frac{1}{d_\ell}\right). \nonumber
\]
Using again expressions $y_i-y_k=(d_i-d_k)/(4\epsilon)$ and relation $d_1+d_2+d_3+d_4=4$, we get:
\[
1-\epsilon^2(y_i+y_j-y_k-y_\ell)^2=1-\frac{1}{16}(d_i+d_j-d_k-d_\ell)^2=\frac{1}{4}(d_i+d_j)(d_k+d_\ell),
\]
Thus, we find:
\beq\nn
\frac{1-\epsilon^2(\t y_i+\t y_j-\t y_k-\t y_\ell)^2}{1-\epsilon^2(y_i+y_j-y_k-y_\ell)^2} =
\frac{(1-\ep s)^2}{S^2}\frac{1}{d_1d_2d_3d_4},
\eeq
which coincides with (\ref{p1}).
\endpf

\section{An alternative discretization of the generalized  Kovalevskaya system}

We now present a discretization of the generalized Kovalevskaya system (\ref{kk3}) which generalizes (\ref{jfg}). It is different from the HK-discretization (\ref{map}), but shares with it $N-2$ integrals of motion and the invariant measure form:
\beq\label{er3r}
\frac{\t y_i}{1-\ep \t y_i}=\frac{y_i}{1+\ep  y_i} \frac{1}{ R(y,\ep)}, \qquad 1 \leq i \leq N,
\eeq
where
$$
R(y,\ep)=1-\ep \sum_{j=1}^N\frac{y_j}{1+\ep y_j}.
$$
This defines a rational map $\t y = f(y, \ep)$. Notice that
\bea
R(y,\ep)&=&\left(\prod_{j=1}^N(1+\epsilon y_j)\right)^{-1}
\left(1-\epsilon\frac{d}{d\epsilon}\right)\prod_{j=1}^N(1+\epsilon y_j)
\nn \\
&=&\left(\prod_{j=1}^N(1+\epsilon y_j)\right)^{-1}
D(y,\epsilon), \nn
\eea
where
$$
D(y,\ep)=1-\sum_{k=2}^{N}  \ep^k (k-1 ) e_{k}(y),
$$
and $e_{k}(y)$ is the elementary symmetric polynomial of degree $k$
in $N$ variables $y$:
$$
e_{k}(y)=\sum_{1\leq j_1 < \dots < j_k \leq N} y_{j_1} \cdots y_{j_k}.
$$
Equations of motion (\ref{er3r}) immediately yield also
\beq\label{op}
\t y_i=\frac{y_i}{1+\ep  y_i}\frac{1}{ R_i(y,\ep)}, \qquad 1 \leq i \leq N,
\eeq
where
\bea
R_i(y,\ep)&=&1-\ep \sum_{j\neq i}\frac{y_j}{1+\ep y_j} \nn \\
&=& \left(\prod_{j\neq i}(1+\epsilon y_j)\right)^{-1}
\left(1-\epsilon\frac{d}{d\epsilon}\right)\prod_{j\neq i}(1+\epsilon y_j)\nn \\
&=& \left(\prod_{j\neq i}(1+\epsilon y_j)\right)^{-1} D_i(y,\epsilon),\nn
\eea
with
$$
D_i(y,\ep)=D(y,\ep)|_{y_i=0}=1-\sum_{k=2}^{N}  \ep^{k} (k-1 ) e_{k}(y_1,\ldots,\cancel{y_i},\ldots,y_N),
$$
where $\cancel{y_i}$ means that the variable $y_i$ is omitted from the list of arguments.

For $N=3$ map (\ref{op}) looks as follows:
$$
\t y_i = y_i \frac{(1+\ep y_j)(1+\ep y_k)}{\left(1+\ep y_i\right)
(1-\ep^2 y_jy_k)},
$$
where $(i,j,k)$ is any permutation of (1,2,3). This is exactly the map (\ref{jfg}) obtained from the
spherical cosine law.

For $N=4$ we get:
$$
\t y_i = y_i \frac{(1+\ep y_j)(1+\ep y_k)(1+\ep y_\ell)}{\left(1+\ep y_i\right)
(1-\ep^2 (y_jy_k+y_j y_\ell+y_k y_\ell)-2\ep^3 y_jy_ky_\ell)},
$$
where $(i,j,k,\ell)$ is any permutation of (1,2,3,4).

It is not immediately clear that the map defined by formula (\ref{er3r}) (or (\ref{op})) is birational. However, this is the case, and can be shown with the help of the following result.

\begin{theorem}\label{TTh8}
For the map (\ref{er3r}), there holds:
$$
R(y,\ep) = \frac{1}{R(\t y,-\ep) }.
$$
\end{theorem}

 \noindent {\it Proof.}
From (\ref{er3r}) it follows that
 $$
R(y,\ep)= 1-\ep \sum_{i=1}^N \frac{y_i}{1+\ep  y_i} = 1 - \ep R(y,\ep)  \sum_{i=1}^N\frac{\t y_i}{1-\ep \t y_i},
 $$
 that is
 $$
 R(y,\ep) \left(1+\ep\sum_{i=1}^N\frac{\t y_i}{1-\ep \t y_i}\right)=1.
 $$
This is the claim of the Theorem.
\endpf

Theorem \ref{TTh8} assures that discrete equations (\ref{er3r}) are equivalent to
\[
\frac{y_i}{1+\ep  y_i}=\frac{\t y_i}{1-\ep \t y_i}\frac{1}{ R(\t y,-\ep)}, \qquad 1 \leq i \leq N,
\]
which is nothing but (\ref{er3r}) with $y\leftrightarrow \t y$ and $\ep\leftrightarrow -\ep$.
This means that the inverse map for $f(y,\ep)$ is given by
$
f^{-1}(y,\ep)=f(y,-\ep),
$
and, in particular, that $f$ is birational.

The following statement shows that the map (\ref{op}) admits the same conserved quantities
and invariant volume form of the HK-discretization (\ref{map}).

\begin{theorem}
${}$
\begin{enumerate}

\item
The functions
\beq \label{444r}
\frac{K_{ij}}{K_{k\ell}}=\frac{y_i-y_j}{y_iy_j}\;\frac{y_ky_\ell}{y_k-y_\ell},
\qquad 1\leq i,j,k,\ell\leq N,
\eeq
are conserved quantities of the map (\ref{op}). Among these, there are $N-2$ functionally independent ones.
\\
\item The map (\ref{op}) admits an invariant volume form
\beq
\omega= \frac{dy_1\wedge \cdots \wedge dy_N}{\psi(y)},\nonumber
\eeq
where $\psi(y)$ is any one of the functions
\beq \nn
\psi(y) = \left( \frac{y_i-y_j}{y_iy_j}\right)^{N-1}\left(\prod_{k=1}^Ny_k\right)^2.
\eeq
\end{enumerate}
\end{theorem}

 \noindent {\it Proof.} (1)
It follows immediately from (\ref{er3r}) that
$$
 \frac{1}{\t y_i}-\frac{1}{\t y_j}= \left(\frac{1}{y_i}-\frac{1}{y_j}\right)R(y,\ep),
$$
so that the quantity
\beq\label{oppp2}
\frac{\t y_i-\t y_j}{\t y_i\t y_j} \frac{y_iy_j}{y_i-y_j}=R(y,\ep)
\eeq
is independent of indices $i,j$. Therefore, functions (\ref{444r}) are integrals of motion.

(2) We have to compute the determinant $\det(\partial\t y/\partial y)$. To this end, we represent the map $y\mapsto\t y$ as a composition of three simpler maps,
$$
y\mapsto u\mapsto v\mapsto\t y,
$$
where
$$
u_i=\frac{y_i}{1+\ep y_i},\qquad
v_i=u_i\left(1-\ep\sum_{j=1}^N u_j \right)^{-1},\qquad \t y_i=\frac{v_i}{1+\ep v_i}.
$$
One easily computes for the first and the third map:
$$
\det \left( \frac{\partial u}{\partial y}\right)=\prod_{k=1}^N\frac{1}{(1+\ep y_k)^2},\qquad
\det \left(\frac{\partial \t y}{\partial v}\right)=\prod_{k=1}^N\frac{1}{(1+\ep v_k)^2}=\prod_{k=1}^N(1-\ep \t y_k)^2.
$$
As for the remaining second map, we compute:
$$
\frac  {\partial v_i}{\partial u_j}=\frac{1}{R^2}(\delta_{ij}R+\ep u_i),\qquad  R=1-\ep\sum_{j=1}^N u_j.
$$
Thus,
\beq
\det \left(\frac{\partial v}{\partial u}\right)=
\det \left( \frac{1}{R^2}(R\, \mathds{1}+\ep ue^{\rm T})\right)
=\frac{1}{R^{N}}\left(1+\frac{1}{R}\ep\langle u,e\rangle\right)=\frac{1}{R^{N+1}}. \nn
\eeq
Collecting everything and using (\ref{er3r}) and (\ref{oppp2}), we find:
\bea
\det \left(\frac{\partial \t y}{\partial y}\right) & = &
\frac{1}{R^{N+1}} \prod_{k=1}^N \frac{(1-\ep \t y_k)^2}{(1+\ep y_k)^2}
=R^{N-1} \prod_{k=1}^N \frac{\t y_k^2}{y_k^2}\nn \\
& = &
\left(\frac{\t y_i-\t y_j}{\t y_i\t y_j}\frac{y_iy_j}{y_i-y_j}\right)^{N-1}
\left(\prod_{k=1}^N \frac{\t y_k}{y_k}\right)^2. \nn
\put(97,0){$\blacksquare$}
\eea

In the case $N=3$, Theorem \ref{THv} tells us that map (\ref{op}) with admits the following conserved quantities:
$$
K_{mn}(\epsilon)=K_{mn}\,\Phi(y,\epsilon),\qquad
\Phi(y,\ep)=\frac{1}{ 1-\epsilon^2 y_iy_j},
$$
where $K_{mn}$ are defined in (\ref{qq3}) and $i,j=1,2,3$, $i \neq j$.

In the case $N=4$, we have the following result.

\begin{theorem}
The map (\ref{op}) with $N=4$ admits the following conserved quantities:
\beq
K_{mn}(\epsilon)=K_{mn}\,\Phi(y,\epsilon),
\qquad
\Phi(y,\ep)=\frac{1}{ \sqrt{(1-\epsilon^2 y_iy_j)(1-\epsilon^2 y_ky_\ell)}}. \label{edef}
\eeq
Here $K_{mn}$ are defined in (\ref{qq4}), and $(i,j,k,\ell)$ is any permutation of $(1,2,3,4)$. Among functions (\ref{edef}), there are  three functionally independent ones.
\end{theorem}

 \noindent {\it Proof.} We have to prove that
\beq \label{8855}
 \left( \frac{\t y_m - \t y_n}{\t y_m \t y_n} \; \frac{y_m y_n}{y_m - y_n} \right)^2
  \frac{ \t y_1 \t y_2 \t y_3 \t y_4}{  y_1 y_2 y_3 y_4 }=
  \frac{(1-\epsilon^2 \t y_i\t y_j)(1-\epsilon^2 \t y_k\t y_\ell)}{(1-\epsilon^2 y_iy_j)(1-\epsilon^2 y_ky_\ell)}.
\eeq
For the l.h.s. of (\ref{8855}), equations (\ref{op}) and (\ref{oppp2}) immediately yield the following representation:
\beq \label{8866}
 \left( \frac{\t y_m - \t y_n}{\t y_m \t y_n} \; \frac{y_m y_n}{y_m - y_n} \right)^2\frac{ \t y_1 \t y_2 \t y_3 \t y_4}{  y_1 y_2 y_3 y_4 }= \frac{D^2(y,\ep)}{D_1(y,\ep)D_2(y,\ep)D_3(y,\ep)D_4(y,\ep)}.
\eeq
Turning to the r.h.s. of (\ref{8855}), we use Eq. (\ref{op}) for $N=4$ to obtain:
$$
1-\epsilon^2 \t y_i\t y_j=1-\frac{\ep^2 y_i y_j (1+ \ep y_k)^2(1+ \ep y_\ell)^2}{D_i(y,\ep)D_j(y,\ep)}.
$$

Now, a direct computation proves the following polynomial identity (valid only if $N=4$):
$$
D_i(y,\ep)D_j(y,\ep)-\ep^2y_i y_j (1+ \ep y_k)^2(1+ \ep y_\ell)^2=(1-\epsilon^2  y_k y_\ell)D(y,\ep).
$$
Therefore, we find:
$$
\frac{1-\epsilon^2 \t y_i\t y_j}{1-\epsilon^2  y_k y_\ell}=\frac{D(y,\ep)}{D_i(y,\ep)D_j(y,\ep)},
$$
so that
\beq \label{8877}
\frac{(1-\epsilon^2 \t y_i\t y_j)(1-\epsilon^2 \t y_k\t y_\ell)}
{(1-\epsilon^2  y_i y_j)(1-\epsilon^2  y_k y_\ell)}=\frac{D^2(y,\ep)}{D_i(y,\ep)D_j(y,\ep)D_k(y,\ep)D_\ell(y,\ep)}.
\eeq
Comparing (\ref{8866}) and (\ref{8877}), we see that (\ref{8855}) holds true.
\endpf

\section{Concluding remarks}
There are several particular problems left open in the above exposition. The most important and difficult one is about the existence or non-existence of the last, $(N-1)$-st, integral of motion of our maps. However, still more important is to understand general reasons for integrability or non-integrability of birational maps generated within the Hirota-Kimura approach. A still more important probram is the study of integrability of birational maps on a possibly general basis. We hope that additional ``experimental'' results presented in this note will be helpful for this ambitious goal.

\section*{Acknowledgements}

We would like to thank Vsevolod Adler for calling our attention to the generalized Kovalevskaya system and for communicating his observation that the cross-ratios of the phase variables are integrals of motion of the HK-discretization of this system.

The authors are partly supported by DFG (Deutsche Forschungsgemeinschaft)in the frame of Sonderforschungsbereich/Transregio 109 ``Discretization in Geometry and Dynamics''.

The participation of the second author in the SIDE 10 conference in Ningbo (China) was financially supported by the organisers, which is gratefully acknowledged.



\end{document}